\documentclass[prl,twocolumn,showpacs,amsmath,amssymb,superscriptaddress]{revtex4}

\usepackage{graphicx}
\usepackage{dcolumn}
\usepackage{bm}
\usepackage{bbm}
\usepackage{ulem}
\usepackage{color}

\begin{document}

\title{Spin Texture and Mirror Chern number in Hg-Based Chalcogenides}

\author{Qingze Wang}
\affiliation{Department of Physics, The Pennsylvania State University, University Park, Pennsylvania 16802-6300, USA}
\author{Shu-chun Wu}
\affiliation{Max Planck Institute for Chemical Physics of Solids, D-01187 Dresden, Germany}
\author{Claudia Felser}
\affiliation{Max Planck Institute for Chemical Physics of Solids, D-01187 Dresden, Germany}
\author{Binghai Yan}\email{yan@cpfs.mpg.de}
\affiliation{Max Planck Institute for Chemical Physics of Solids, D-01187 Dresden, Germany}
\affiliation{Max Planck Institute for Physics of Complex Systems, D-01187 Dresden, Germany}
\author{Chao-xing Liu}\email{cxl56@psu.edu}
\affiliation{Department of Physics, The Pennsylvania State University, University Park, Pennsylvania 16802-6300, USA}

\date{\today}

\begin{abstract}
The unique feature of surface states in topological insulators is the so-called ``spin-momentum locking'', which means that electron spin is oriented along a fixed direction for a given momentum and forms a texture in the momentum space. In this work, we study spin textures of two typical topological insulators in Hg-Based Chalcogenides, namely HgTe and HgS, based on both the first principles calculation and the eight band Kane model. We find opposite helicities of spin textures between these two materials, originating from the opposite signs of spin-orbit couplings. Furthermore, we reveal that different mirror Chern numbers between HgTe and HgS characterize different topological natures of the systems with opposite spin textures and guarantee the existence of gapless interface states.
\end{abstract}
\maketitle

{\it Introduction -}
A recent discovery in condensed matter physics is the theoretical prediction and experimental observation of time reversal (TR) invariant topological insulators (TIs)\cite{moore2009,hasan2010,qi2010,qi2011}. TR invariant TIs are insulating in the bulk, similar to ordinary insulators. However, their surfaces possess conducting channels, of which the gapless nature is protected by TR symmetry. The topological surface states (TSSs) are unique in the sense that their spin direction is locked to the momentum, forming a helical structure in the momentum space. Thus, the TSSs of TIs are also dubbed ``helical metals''\cite{wu2006}. The helicity of spin texture is usually left-handed for various materials, such as Bi$_2$Se$_3$ family of materials\cite{zhangh2009}. It has been shown that the left-handed spin textures in the Bi$_2$Se$_3$ family can be directly related to the atomic spin-orbit coupling (SOC) \cite{liuc2010,zhangh2013}. Therefore, it is natural to ask if the relationship between spin textures and SOC is general or not. In addition, it has recently been shown that mirror Chern numbers\cite{teo2008,hsieh2012} can distinguish different topological natures of TIs with opposite spin textures, leading to topologically protected interface states and unusual transport behaviors\cite{takahashi2011}. Thus, it is desirable to find realistic TI materials with right-handed spin textures.

In this letter, we investigate spin textures, as well as mirror Chern numbers, of topological states in Hg based Chalcogenides, mainly focusing on HgTe and HgS. In particular, based on the first-principles calculation, we find that although both materials are TIs with a single Dirac cone at one surface, the helicities of spin textures are opposite between HgTe and HgS. The physical origin is due to the opposite signs of SOC in these two materials, which can be understood from the eight band Kane model. We show that mirror Chern numbers are different in these two materials and distinguish them into different topological phases. Furthermore, we confirm their different topological natures by directly calculating topologically non-trivial gapless states at the interface between HgS and HgTe. The relationship between mirror Chern number and spin textures allows us to utilize spin textures as a simple and natural approach to distinguish topological phases with different mirror Chern numbers in the first-principles calculation.

\begin{figure}[tb]
	\includegraphics[width = 0.9\columnwidth,angle=0]{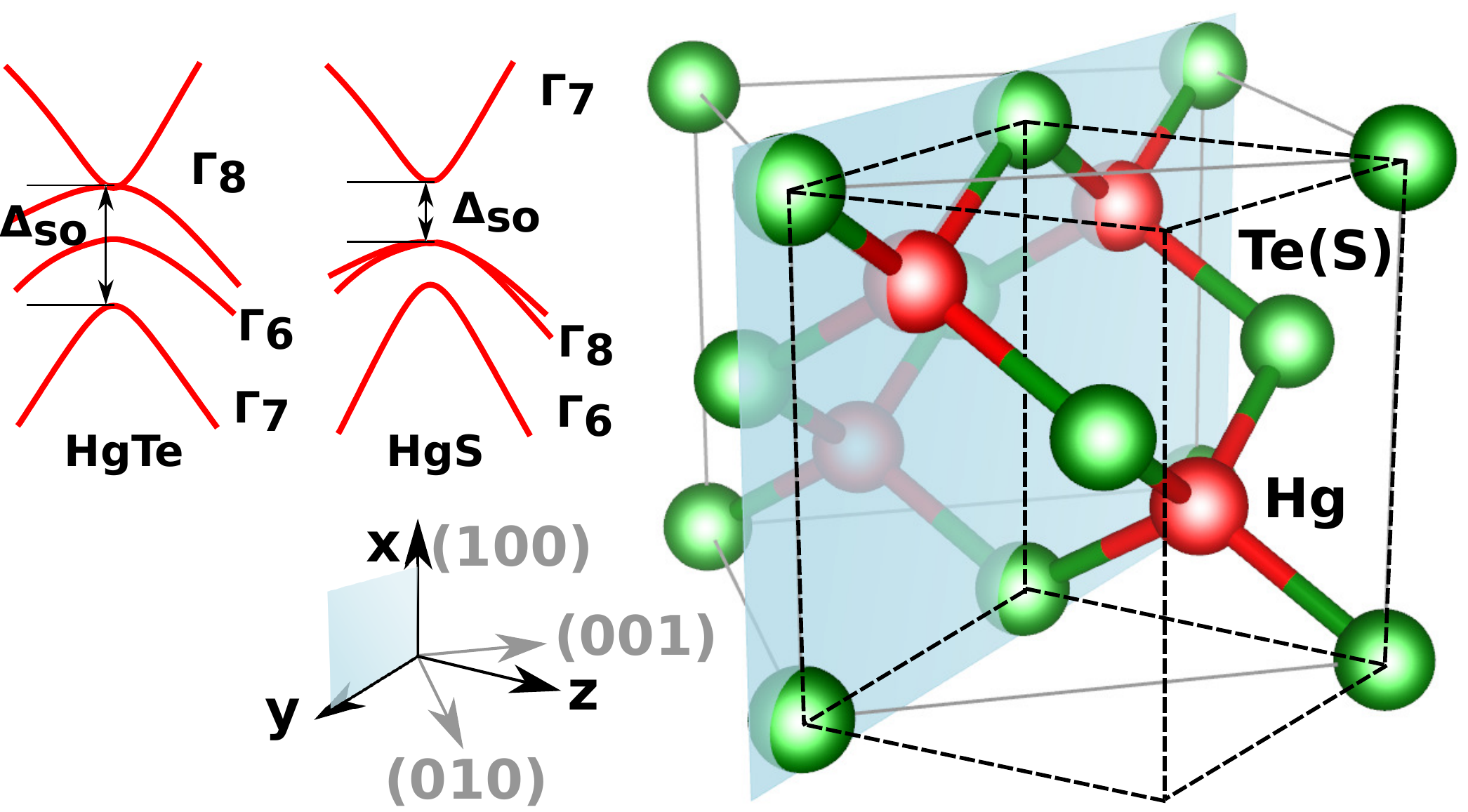}
	\caption{(Color online) Zinc-blende crystal structures and electronic structure for HgTe and HgS. The blue plane indicates the mirror plane for the $(011)$ mirror symmetry. The band alignment of $\Gamma_8$ bands and $\Gamma_7$ bands is opposite for HgTe and HgS due to the opposite sign of the SOC parameter $\Delta_{so}$.  }
    \label{fig1}
\end{figure}

\begin{figure}[tb]
	\includegraphics[width = 1.0\columnwidth,angle=0]{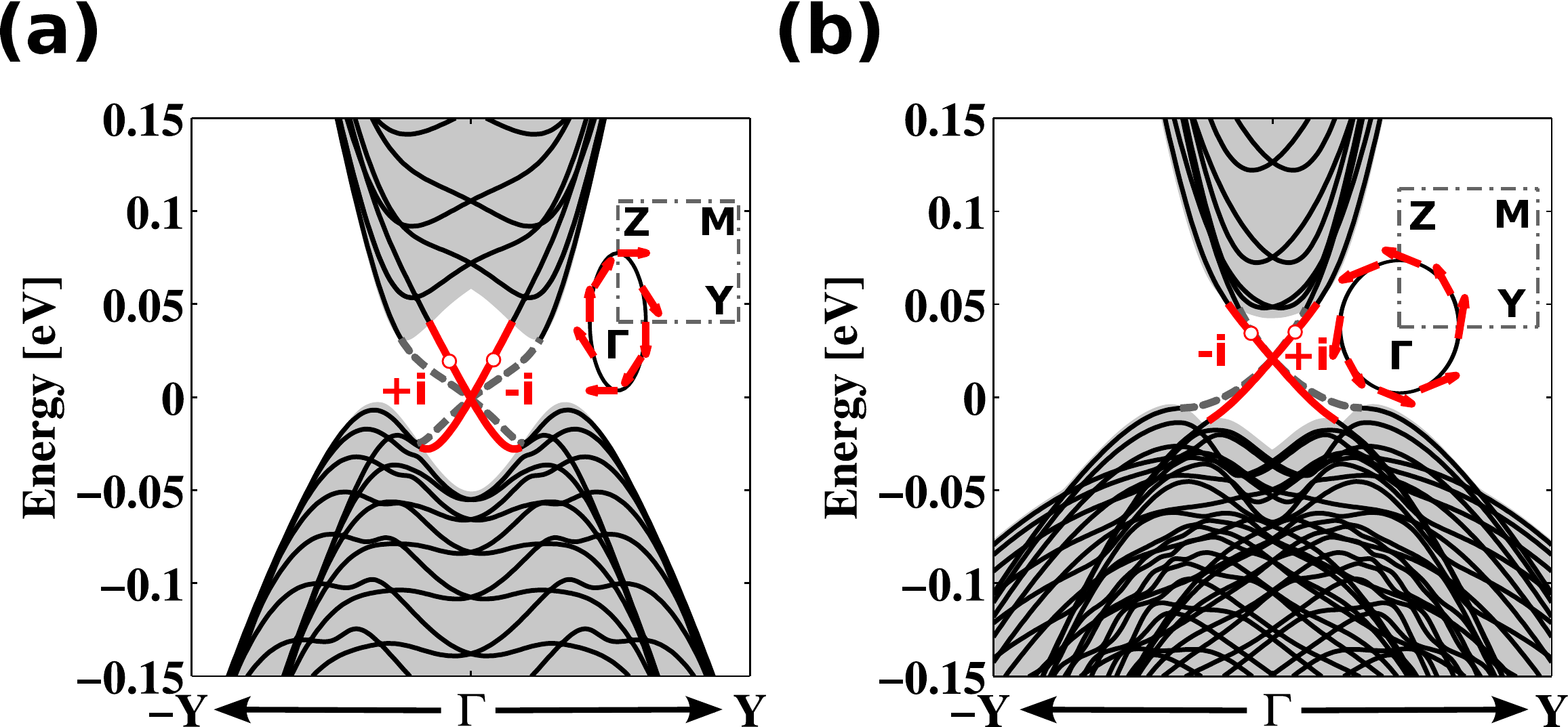}
  \caption{(color online) Band structures for (a) HgTe and (b) HgS by first-principles calculation. The surface Brilloun zone is the $k_y - k_z$ plane and the band structure is shown along one third of $\Gamma- -Y$ and $\Gamma-Y$ lines.
   Solid red and gray dashed lines indicate the top and bottom surface states, respectively. The band alignment of the bulk band structure and the spin texture of the upper surface Dirac cone on the top surface are illustrated on the left and right insets, respectively.  For HgTe, 5\% uniaxial strain is applied to open the bulk energy gap. The gray shadow region represents the projection of bulk states.
   }
    \label{fig2}
\end{figure}

{\it Spin textures for HgTe and HgS -}
HgTe and HgS are II-VI compound semiconductors with zinc-blende crystal structures, as shown in Fig. \ref{fig1}. Their band structures exhibit a band inversion at the $\Gamma$ point, which induces the TI phase in both compounds \cite{bernevig2006,moon2006,svane2011,virot2013}. The band inversion occurs between Hg-$s$ and Te-$p$ (S-$p$) bands for HgTe (HgS) near the Fermi energy. According to the symmetry of the wave function, Hg-$s$ bands are labelled as $\Gamma_6$ bands and Te-$p$ or S-$p$ bands split into $\Gamma_8$ ($j=3/2$) and $\Gamma_7$ ($j=1/2$) bands due to the SOC. The first-principles calculation shows that $s$-orbital-like $\Gamma_6$ bands lies below $p$-orbital-like $\Gamma_8$ ($\Gamma_7$) bands in HgTe (HgS), leading to the inverted band structures. The sequence of $\Gamma_8$ and $\Gamma_7$ bands, which is determined by the so-called SOC splitting, as denoted by $\Delta_{so}$ in Fig. ~\ref{fig1}, are opposite for these two materials. HgTe exhibits a normal SOC splitting, i.e. $\Gamma_8$ ($j=3/2$) bands are above $\Gamma_7$ ($j=1/2$), so that light-hole and heavy-hole bands of $\Gamma_8$ states form the lowest conduction band and highest valence band, respectively. Because of the cubic symmetry, the light hole and heavy hole bands are degenerate at the $\Gamma$ point, yielding a zero energy gap in HgTe (semi-metal phase). In contrast, HgS shows a ``negative'' SOC splitting \cite{carrier2004,virot2013}, in which $\Gamma_8$ bands are below $\Gamma_7$ band. Thus, the valence and conduction bands are formed by $\Gamma_8$ and $\Gamma_7$ bands, respectively, with a non-zero energy gap. In order to reveal TSSs for both compounds, we performed density-functional theory band structure calculations ~within the local-density approximation (LDA) framework, which is implemented in the
Vienna \textit{Ab-initio} Simulation Package (\textsc{vasp})~\cite{kresse1996}. Surfaces were simulated in a slab model with the surface normal along the $x$ direction based on maximally localized Wannier functions~\cite{mostofi2008} that extracted from the first-principles calculation on bulk materials. The top and down surfaces of the slab were terminated by Te (S) atomic layers for HgTe (HgS) as a boundary condition, which accommodates simple surface states (see below). In order to open an energy gap of HgTe, we applied a compressive strain of 5\% along the $x$ axis. The calculated surface band structures are shown in Fig. 1. For both HgTe and HgS, gapless surface states with linear dispersions that form a single Dirac cone in the surface Brilloun zone exist on the top or bottom surface for both materials, as seen in Fig.\ref{fig2} (a) and (b). The TSSs on the top and bottom surfaces show different dispersions because of the lack of inversion symmetry in the zinc-blende lattice. From the Fermi surface plot, one can find that the TSSs of HgTe are anisotropic in the surface $k_y-k_z$ plane, while those of HgS are relatively isotropic. The spin of the TSSs mainly lies in the surface plane while a small amount of the $x$-direction spin is also found. The in-plane spin components of the TSSs form a texture, as shown on the Fermi surface in Fig. \ref{fig2}. We find that the upper Dirac cone of the top surface displays a left-handed spin texture for HgTe, similar to that in Bi$_2$Se$_3$ family of materials, while the spin texture of HgS is right-handed. Since the essential difference between HgTe and HgS is the opposite SOC splitting, it is natural to expect that the opposite spin textures are related to SOC, which will be analyzed in details below.

\begin{figure}[tb]
    \begin{center}
	    \includegraphics[width=1.05\columnwidth,angle=0]{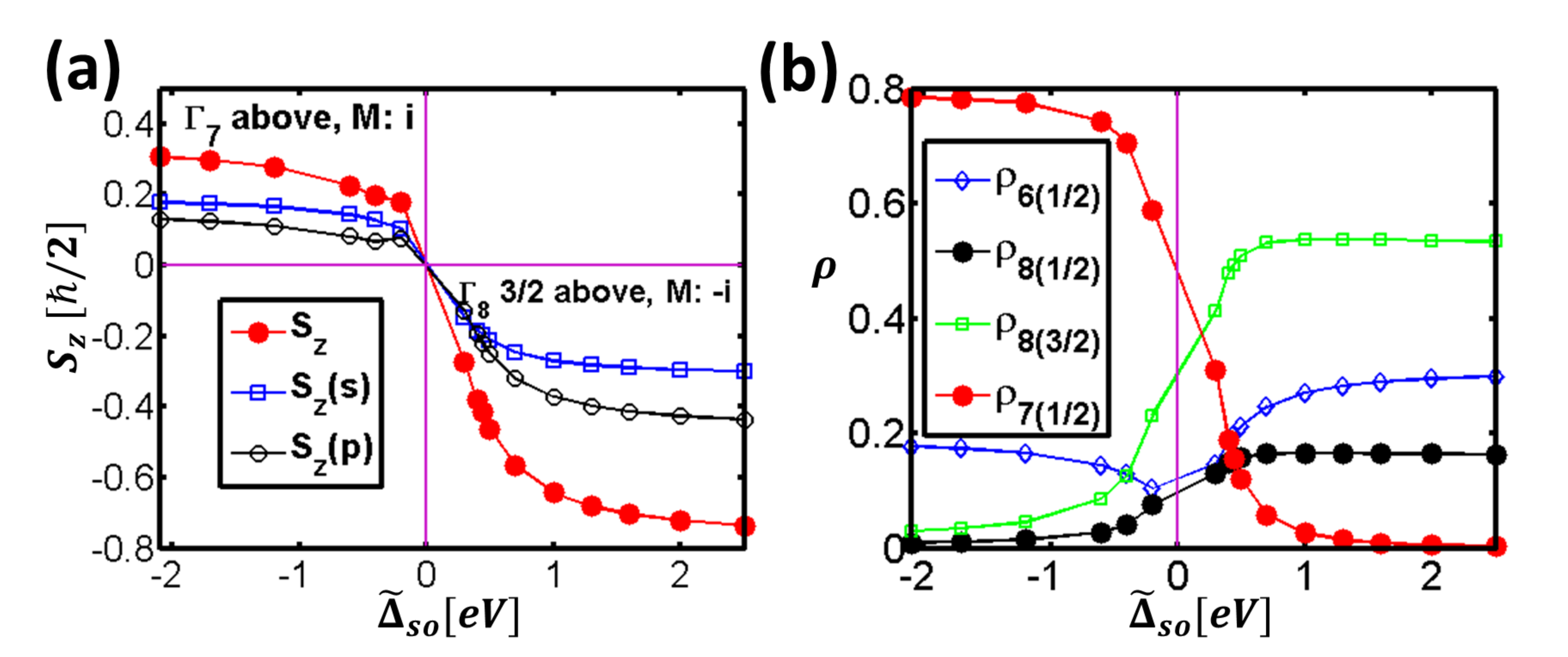}
    \end{center}
 \caption{
 (Color online) (a) The expectation value of the z-direction spin ($S_z$) at the momentum ($k_y= 0.02$\AA$^{-1}$) is plotted as a function of the SOC parameter $\tilde{\Delta}_{so}$ for the top surface of the slab along the x direction. The red dots, blue squares and black circles represent total spin, s orbital contributed spin and p orbital contributed spin, respectively. (b) Wave function components of TSSs at the momentum ($k_y= 0.02$\AA$^{-1}$) are plotted as a function of $\tilde{\Delta}_{so}$ for the top surface. Different colors are for different wave function components. All the calculations in this figure are done with the Kane model.
 }
    \label{fig3}
\end{figure}

Since all the relevant physics occurs near the $\Gamma$ point for both HgTe and HgS, we can adopt the eight band Kane model, which is a standard model to describe II-VI and III-V semiconductors with the zinc-blende or diamond structures\cite{winkler2003,novik2005}, to analyze spin textures in these systems. The details of the Kane model are described in the appendix\cite{appendix}. We calculate the energy dispersion of HgS and HgTe in the slab configuration along the x direction with the Kane model. TSSs with a single Dirac cone are found inside the band gap for both HgS and HgTe. Furthermore, we numerically calculate spin textures of TSSs, which are left-handed for HgTe and right-handed for HgS. All the results are in good agreement with those from first-principles calculations. As described above, the main difference between HgTe and HgS is the opposite alignment between the $\Gamma_8$ and $\Gamma_7$ bands. In the Kane model, the band sequence is controlled by the SOC parameters $\Delta_{so}$. To understand the relationship between spin textures and SOC, we check the energy dispersion and spin textures by tuning the SOC parameter $\Delta_{so}$, but keeping all the other parameters fixed. We have checked the spin orientation at the momentum ${\bf k}_0=(0,0.02,0)$\AA$^{-1}$ as a function of $\tilde{\Delta}_{so}$, as shown in the Fig. \ref{fig3}(a). Here we have re-define SOC parameters as $\tilde{\Delta}_{so}= \Delta_{so} + a \epsilon_{xx}$ where $a\epsilon_{xx} $ describes the energy shift of the $\Gamma_8$ bands due to the x-direction strain \cite{novik2005}. $\tilde{\Delta}_{so}$ reflects the energy difference between $|\Gamma_8,\pm \frac{3}{2}\rangle$ bands and $|\Gamma_7,\pm\frac{1}{2}\rangle$ bands.
It is clear that the spin orientation changes its direction when the band sequence of $|\Gamma_8,\pm \frac{3}{2}\rangle$ and $|\Gamma_7,\pm\frac{1}{2}\rangle$ bands is reversed. This calculation confirms that the helicity of spin texture is determined by the SOC parameter $\Delta_{so}$. In addition, we decompose the wave functions of surface states into the basis of the eight band Kane model and observe that the main contributions to TSSs near $\Gamma$ point are from $|\Gamma_6, \pm\frac{1}{2}\rangle$ and $|\Gamma_8,\pm\frac{3}{2}\rangle$ ($|\Gamma_7,\pm\frac{1}{2}\rangle$) bands for HgTe(HgS), shown in Fig. \ref{fig3}(b).
One may notice that the $|\Gamma_8,\pm\frac{1}{2}\rangle$ band also contributes to surface states in HgTe. However, the $|\Gamma_8,\pm\frac{1}{2}\rangle$ band is occupied for both HgTe and HgS. Thus, this band is not important for topological distinction between these two materials. Therefore, we will study an effective four band model with the basis $|\Gamma^6,\pm\frac{1}{2}\rangle$ and $|\Gamma^8,\pm\frac{3}{2}\rangle$ ($|\Gamma^7,\pm\frac{1}{2}\rangle$) for HgTe (HgS) below.

{\it Mirror Chern number and Topological interface states -}
Next we would like to study more carefully how the SOC determines spin helicities of TSSs in TIs. In addition, one may ask if the difference in spin textures has any topological meaning. Below we will show that these two materials are topologically distinguished by opposite mirror Chern numbers. Thus, a robust gapless state exists at the interface between HgTe and HgS.

\begin{figure*}[tb]
    \includegraphics[width = 1.9\columnwidth,angle=0]{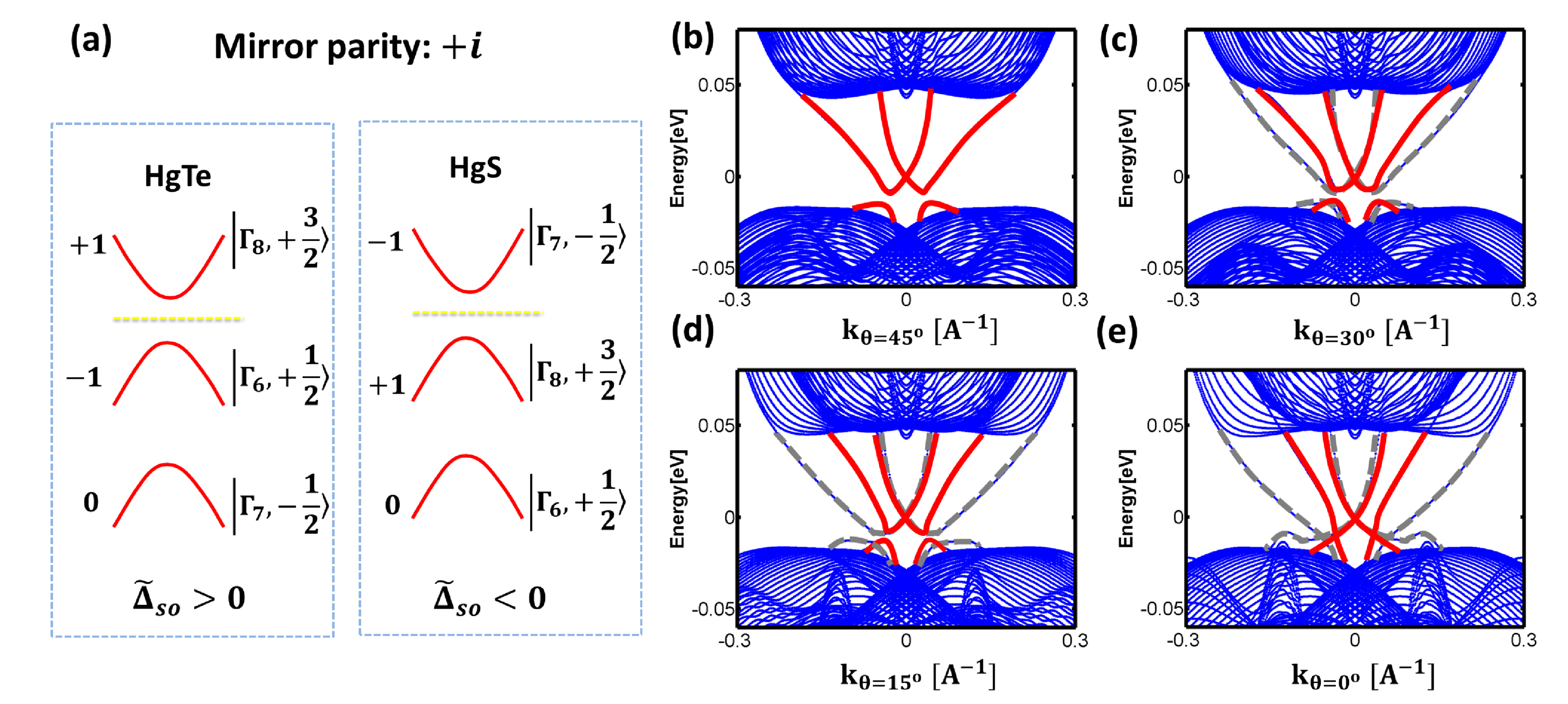}
  \caption{(Color online) (a) Schematic of bulk band structure of HgTe and HgS in mirror parity $+i$ subspace. The yellow dashed lines denote Fermi
  energy. +1, -1 and 0 represent the Chern number for each band. (b-e), Mirror symmetry protected interface gapless states for HgTe/HgS superlattice along different directions, $\theta = 45^o$, $\theta = 30^o$, $\theta = 15^o$ and $\theta = 0^o$, respectively. Here $\theta = tan^{-1}(\frac{k_y}{k_z})$. Solid red and gray dashed lines indicate the top and bottom interfacial states, respectively. To make the gapless nature of interfacial states clear, we adjust some parameters for the Kane model in our calculation, which should not affect topological natures.  }
    \label{fig4}
\end{figure*}

The zinc-blende crystal structure of HgTe or HgS has the mirror symmetry with respect to ({$011$}) and ($01\bar{1}$) planes, which is preserved for the slab along the [100] direction, as illustrated in Fig. \ref{fig1}. Correspondingly, the Hamiltonian of the Kane model is invariant under these symmetry operations. To simplify our analysis, we take the isotropic approximation for the Kane model and choose [$011$] and [$01\bar{1}$] as the z and y axis in Fig. \ref{fig1}, respectively. In this coordinate system, the Kane model is invariant under the mirror operation $\mathcal{M}_{z}$ along the $z$ direction ($\mathcal{M}_{z} H_{Kane}(k_x, k_y, k_z) \mathcal{M}^{-1}_{z} = H_{Kane}(k_x, k_y, -k_z)$). Therefore, at the $k_z=0$ plane, $H_{Kane}$ commutes with $\mathcal{M}_z$ and all the eigen states can be classified by mirror parity of $\mathcal{M}_z$, which takes the value of $\pm i$ after taking into account spin. Furthermore, one can easily show that the basis of the Kane model also have definite mirror parity of $\mathcal{M}_z$. Correspondingly, the Hamiltonian is block diagonal at the $k_z=0$ plane, with each block labelled by mirror parity. Since TR operation can change mirror parity from $+i$ to $-i$, these two blocks can be related to each other by TR symmetry. Therefore, we only need to investigate one block. Let us consider topological properties of the block with the mirror parity $+i$, which is consisted of the basis $|\Gamma_6,\frac{1}{2}\rangle$, $|\Gamma_8,\frac{3}{2}\rangle$, $|\Gamma_8,-\frac{1}{2}\rangle$ and $|\Gamma_7,-\frac{1}{2}\rangle$\cite{appendix}.
As shown in Fig. \ref{fig4}(a), for the mirror parity $+i$, the conduction band of HgTe is consisted of the $|\Gamma_8,\frac{3}{2}\rangle$ band while that of HgS consisted of the $|\Gamma_7,-\frac{1}{2}\rangle$ band. (Here the $|\Gamma_8,\frac{1}{2}\rangle$ band is neglected for the reason mentioned above. ) This difference turns out to result in different mirror Chern numbers in HgTe and HgS. To see this, we may consider the low energy effective theory.
For HgTe, we focus on the basis $|\Gamma_6, \frac{1}{2}\rangle$ and $|\Gamma_8,\frac{3}{2}\rangle$ bands and the corresponding Hamiltonian is given by \begin{eqnarray}
  &&H_{+i,HgTe}=\left(
	\begin{array}{cc}
          T&-\frac{1}{\sqrt{2}}Pk_+ \\
          -{\frac{1}{\sqrt2}}Pk_-&U-V+a\epsilon_{xx}\\
    \end{array}
\label{eq:2-HgTe}
	\right)
\end{eqnarray}
where the expressions $U$, $T$, $V$, $P$ and $a\epsilon_{xx}$ are defined in the appendix\cite{appendix}.
It should be noted that the upper off-diagonal part of the Hamiltonian is proportional to $k_+$, originating from the fact that the total angular momentum of the $|\Gamma_8,\frac{3}{2}\rangle$ band is larger by 1 than that of the $|\Gamma_6,\frac{1}{2}\rangle$ state. This two band model, which has been well studied in Ref. \onlinecite{qi2006}, is topologically non-trivial with the Chern number -1. Once the Chern number for the block with mirror parity $+i$ is obtained, the mirror Chern number, defined as $n_M = (n_{+i} - n_{-i})/2$, can be easily calculated as $-1$. Therefore, we conclude that HgTe is not only a TI, but also a mirror Chern insulator with the mirror Chern number $n_M = -1$. For HgS, we consider the effective model consisted of $|\Gamma_6,\frac{1}{2}\rangle$ and $|\Gamma_7,-\frac{1}{2}\rangle$ bands. We notice that the $|\Gamma_8,\frac{3}{2}\rangle$ band is above the $|\Gamma_6,\frac{1}{2}\rangle$ band and more close to energy gap in realistic HgS (See Fig. \ref{fig1} or \ref{fig4}(a)). However, since both bands are occupied, the interchange between them has no influence to topological properties of the system. The effective Hamiltonian is written as
\begin{eqnarray}
  &&H_{+i,HgS}=\left(
	\begin{array}{cc}
          T&-\frac{1}{\sqrt{3}} Pk_-\\
          -\frac{1}{\sqrt{3}}Pk_+&U-\Delta_{so}
    \end{array}
\label{eq:2-HgS}
	\right).
\end{eqnarray}
Different from the case in HgTe, the upper off-diagonal term for this Hamiltonian is proportional to $k_-$ because the total angular momentum of the $|\Gamma_7,-\frac{1}{2}\rangle$ band is smaller by 1 than that of $|\Gamma_6,\frac{1}{2}\rangle$. Correspondingly, the Chern number in this case is found to be +1, leading to the mirror Chern number $n_M = +1$ for HgS. Therefore, we conclude that the mirror Chern number for HgS is opposite to that for HgTe. The above argument also suggests that one can identify the mirror Chern number by looking at the difference of total angular momenta between two inverted bands with the mirror parity $+i$.

The above discussions have explicitly shown that HgTe and HgS have opposite helicities of spin textures, as well as opposite mirror Chern numbers. This naturally leads to the question: is there any relationship between spin textures and mirror Chern numbers in these two systems? Since the mirror symmetry $\mathcal{M}_{z}$ is also preserved for a slab configuration, we may study the mirror properties of TSSs at $k_z=0$. To simplify our discussion, we again consider the effective theory with the basis $|\Gamma_6, \pm \frac{1}{2}\rangle$ and $|\Gamma_8, \pm \frac{3}{2}\rangle$ ($|\Gamma_7, \pm \frac{1}{2}\rangle$) for HgTe (HgS). In the subspace with the mirror parity $+i$, the TSSs for HgTe (HgS) can be written in the form $|\Psi_{+i}\rangle = \alpha_1 |\Gamma_6,\frac{1}{2}\rangle + \alpha_2 \vert \Gamma_8, \frac{3}{2}\rangle (\vert \Gamma_7, -\frac{1}{2}\rangle)$, where $\alpha_i$ are the coefficients of different orbital components. Correspondingly, the expectation value of the z-direction spin operator $S_z = \frac{\hbar}{2}\sigma_z$ can be calculated as $\langle\Psi_{+i}| S_z |\Psi_{+i}\rangle = \frac{\hbar}{2}|\alpha_1|^2 + \frac{\hbar}{2} |\alpha_2|^2(\frac{\hbar}{2} \frac{1}{3}|\alpha_2|^2)$ for HgTe(HgS). Similar calculation can be applied to the subspace with the mirror parity $-i$, leading to $\langle\Psi_{-i}| S_z |\Psi_{-i}\rangle = -\frac{\hbar}{2}|\alpha_1|^2 - \frac{\hbar}{2} |\alpha_2|^2(-\frac{\hbar}{2} \frac{1}{3}|\alpha_2|^2)$, where $|\Psi_{-i}\rangle = \alpha_1 |\Gamma_6,-\frac{1}{2}\rangle + \alpha_2 \vert \Gamma_8, -\frac{3}{2}\rangle (\vert \Gamma_7, \frac{1}{2}\rangle)$ for HgTe(HgS) (Here we ignore $|\Gamma_8, \pm\frac{1}{2} \rangle$ bands because of their negligible contribution to the TSSs as mentioned above.). Therefore, this calculation demonstrates that the spin orientation follows the mirror parity for both HgTe and HgS. Once the relationship between spin orientation and mirror parity is established, the helicity of spin textures can be easily shown to be related to mirror Chern number. Let us take the mirror parity +i subspace as an example, in which spin is oriented along the positive z direction for both HgTe and HgS. However, in this subspace, the Chern number, which determines the velocity direction of gapless edge states at the boundary of the x-y mirror plane, takes $-1$($+1$) for HgTe(HgS). Correspondingly, the up spin state is a left(right) mover for HgTe(HgS) along y direction, which represents a left-handed(right-handed) spin texture for HgTe(HgS). Thus, the helicity of spin texture can be determined by mirror Chern number for both HgTe and HgS.

A direct physical consequence of opposite mirror Chern numbers in HgTe and HgS is the existence of robust gapless states\cite{takahashi2011} at the interface between HgTe and HgS once mirror symmetry is preserved in the hetero-structure. To see this, we perform a numerical calculation of a superlattice structure made of HgTe and HgS based on the Kane model. The complexity of the interface is neglected. As depicted in Fig. \ref{fig4}(b-e), the gapless interface states only survive along the direction ${\bf k}= (0,k_y,0)$ and ${\bf k}= (0,0,k_z)$ (Fig. \ref{fig4}(e)), which correspond to the mirror symmetric planes ($01\bar{1}$) and ($011$), while interface states are gaped along other directions due to the breaking of mirror symmetry (Fig. \ref{fig4}(b-d)).

{\it Conclusion and Discussion - }
In conclusion, we have shown opposite spin textures of TSSs and also opposite mirror Chern numbers between HgTe (with appropriate strain) and HgS. Experimentally, one could use spin-resolved angle-resolved photoemission spectroscopy (ARPES) \cite{hsieh2009a,hsieh2009b,souma2011,xu2011,pan2011,jozwiak2011} to test spin textures for HgTe and HgS. It has also been proposed that the opposite mirror Chern numbers will lead to a unique feature in the reflectance and transmittance with normal incident electrons\cite{takahashi2011}.
One remaining question is that why HgS has opposite sign of the SOC parameter $\Delta_{so}$. For the pure $p$-orbital, one can easily show that with the atomic SOC, the $S=\frac{3}{2}$ states (such as $\Gamma_8$ bands) are always above the $S=\frac{1}{2}$ states (such as $\Gamma_7$ bands). The unique feature of HgS is the strong hybridization between the S-$3p$ and Hg-$5d$ orbitals near the Fermi surface\cite{cardona1963,shindo1965,zunger2012}. Due to the low symmetry of zinc-blende structures, $t2g$ states of Hg-$5d$ orbitals can contribute a large amount to the $\Gamma_7$ state through the $p-d$ hybridization. As shown in Ref. \onlinecite{shindo1965}, since the $t2g$ states originate from the $S=\frac{5}{2}$ states, the corresponding $\Gamma_7$ states are always above the $\Gamma_8$ states for $d$-orbitals, opposite to the case of $p$-orbitals. This leads to the negative SOC parameters in HgS. This analysis suggests to search for right-handed topological materials in the TI materials with d-orbitals near the Fermi energy. One can find HgS analogs among Heulser materials~\cite{chadov2010,lin2010,xiao2010}, which can be treated as ternary counterparts to zinc-blende compounds. For example, LiAuS-type of half Heusler compounds were reported to be TIs with negative SOC~\cite{zunger2012} and indeed found to exhibit the right hand spin texture~\cite{lin2014}. Additionally, the fact that many $d$-orbital based compounds such as pyrochlores were discovered as TIs or topological Mott insulators~\cite{Pesin2010,guo2009,Wan2011} indicates possibly the existence of unusual spin textures in such systems.

We would like to thank Xin Liu, C. Br{\"u}ne, C. Ortix and J. van den Brink for useful discussions. This work is supported by  ERC Advanced Grant (291472).

\bibliography{HgTeHgS}

\begin{thebibliography}{10}%
\makeatletter
\providecommand \@ifxundefined [1]{%
 \ifx #1\undefined \expandafter \@firstoftwo
 \else \expandafter \@secondoftwo
\fi
}%
\providecommand \@ifnum [1]{%
 \ifnum #1\expandafter \@firstoftwo
 \else \expandafter \@secondoftwo
\fi
}%
\providecommand \enquote [1]{``#1''}%
\providecommand \bibnamefont  [1]{#1}%
\providecommand \bibfnamefont [1]{#1}%
\providecommand \citenamefont [1]{#1}%
\providecommand\href[0]{\@sanitize\@href}%
\providecommand\@href[1]{\endgroup\@@startlink{#1}\endgroup\@@href}%
\providecommand\@@href[1]{#1\@@endlink}%
\providecommand \@sanitize [0]{\begingroup\catcode`\&12\catcode`\#12\relax}%
\@ifxundefined \pdfoutput {\@firstoftwo}{%
 \@ifnum{\z@=\pdfoutput}{\@firstoftwo}{\@secondoftwo}%
}{%
 \providecommand\@@startlink[1]{\leavevmode\special{html:<a href="#1">}}%
 \providecommand\@@endlink[0]{\special{html:</a>}}%
}{%
 \providecommand\@@startlink[1]{%
  \leavevmode
  \pdfstartlink
   attr{/Border[0 0 1 ]/H/I/C[0 1 1]}%
   user{/Subtype/Link/A<</Type/Action/S/URI/URI(#1)>>}%
  \relax
 }%
 \providecommand\@@endlink[0]{\pdfendlink}%
}%
\providecommand \url  [0]{\begingroup\@sanitize \@url }%
\providecommand \@url [1]{\endgroup\@href {#1}{\urlprefix}}%
\providecommand \urlprefix [0]{URL }%
\providecommand \Eprint[0]{\href }%
\@ifxundefined \urlstyle {%
  \providecommand \doi [1]{doi:\discretionary{}{}{}#1}%
}{%
  \providecommand \doi [0]{doi:\discretionary{}{}{}\begingroup
  \urlstyle{rm}\Url }%
}%
\providecommand \doibase [0]{http://dx.doi.org/}%
\providecommand \Doi[1]{\href{\doibase#1}}%
\providecommand \bibAnnote [3]{%
  \BibitemShut{#1}%
  \begin{quotation}\noindent
    \textsc{Key:}\ #2\\\textsc{Annotation:}\ #3%
  \end{quotation}%
}%
\providecommand \bibAnnoteFile [2]{%
  \IfFileExists{#2}{\bibAnnote {#1} {#2} {\input{#2}}}{}%
}%
\providecommand \typeout [0]{\immediate \write \m@ne }%
\providecommand \selectlanguage [0]{\@gobble}%
\providecommand \bibinfo [0]{\@secondoftwo}%
\providecommand \bibfield [0]{\@secondoftwo}%
\providecommand \translation [1]{[#1]}%
\providecommand \BibitemOpen[0]{}%
\providecommand \bibitemStop [0]{}%
\providecommand \bibitemNoStop [0]{.\EOS\space}%
\providecommand \EOS [0]{\spacefactor3000\relax}%
\providecommand \BibitemShut [1]{\csname bibitem#1\endcsname}%
\bibitem{moore2009}%
  \BibitemOpen
  \bibfield{author}{%
  \bibinfo {author} {\bibfnamefont{J.}~\bibnamefont{Moore}},\ }%
  \bibfield{journal}{%
  \bibinfo {journal} {Nature Physics}\ }%
  \textbf{\bibinfo {volume} {5}},\ \bibinfo {pages} {378} (\bibinfo {year}
  {2009})%
  \bibAnnoteFile{NoStop}{moore2009}%
\bibitem{hasan2010}%
  \BibitemOpen
  \bibfield{author}{%
  \bibinfo {author} {\bibfnamefont{M.~Z.}\ \bibnamefont{Hasan}}\ and\ \bibinfo
  {author} {\bibfnamefont{C.~L.}\ \bibnamefont{Kane}},\ }%
  \bibfield{journal}{%
  \Doi{10.1103/RevModPhys.82.3045}{\bibinfo {journal} {Rev. Mod. Phys.}}\ }%
  \textbf{\bibinfo {volume} {82}},\ \bibinfo {pages} {3045} (\bibinfo {year}
  {2010})%
  \bibAnnoteFile{NoStop}{hasan2010}%
\bibitem{qi2010}%
  \BibitemOpen
  \bibfield{author}{%
  \bibinfo {author} {\bibfnamefont{X.-L.}\ \bibnamefont{Qi}}\ and\ \bibinfo
  {author} {\bibfnamefont{S.-C.}\ \bibnamefont{Zhang}},\ }%
  \bibfield{journal}{%
  \bibinfo {journal} {Physics Today}\ }%
  \textbf{\bibinfo {volume} {63}},\ \bibinfo {pages} {33} (\bibinfo {year}
  {2010})%
  \bibAnnoteFile{NoStop}{qi2010}%
\bibitem{qi2011}%
  \BibitemOpen
  \bibfield{author}{%
  \bibinfo {author} {\bibfnamefont{X.-L.}\ \bibnamefont{Qi}}\ and\ \bibinfo
  {author} {\bibfnamefont{S.-C.}\ \bibnamefont{Zhang}},\ }%
  \bibfield{journal}{%
  \bibinfo {journal} {Rev. Mod. Phys.}\ }%
  \textbf{\bibinfo {volume} {83}},\ \bibinfo {pages} {1057} (\bibinfo {year}
  {2011})%
  \bibAnnoteFile{NoStop}{qi2011}%
\bibitem{wu2006}%
  \BibitemOpen
  \bibfield{author}{%
  \bibinfo {author} {\bibfnamefont{C.}~\bibnamefont{Wu}}, \bibinfo {author}
  {\bibfnamefont{B.~A.}\ \bibnamefont{Bernevig}},\ and\ \bibinfo {author}
  {\bibfnamefont{S.-C.}\ \bibnamefont{Zhang}},\ }%
  \bibfield{journal}{%
  \bibinfo {journal} {Phys. Rev. Lett.}\ }%
  \textbf{\bibinfo {volume} {96}},\ \bibinfo {pages} {106401} (\bibinfo {month}
  {Mar}\ \bibinfo {year} {2006})%
  \bibAnnoteFile{NoStop}{wu2006}%
\bibitem{zhangh2009}%
  \BibitemOpen
  \bibfield{author}{%
  \bibinfo {author} {\bibfnamefont{H.}~\bibnamefont{Zhang}}, \bibinfo {author}
  {\bibfnamefont{C.-X.}\ \bibnamefont{Liu}}, \bibinfo {author}
  {\bibfnamefont{X.-L.}\ \bibnamefont{Qi}}, \bibinfo {author}
  {\bibfnamefont{X.}~\bibnamefont{Dai}}, \bibinfo {author}
  {\bibfnamefont{Z.}~\bibnamefont{Fang}},\ and\ \bibinfo {author}
  {\bibfnamefont{S.-C.}\ \bibnamefont{Zhang}},\ }%
  \bibfield{journal}{%
  \bibinfo {journal} {Nature physics}\ }%
  \textbf{\bibinfo {volume} {5}},\ \bibinfo {pages} {438} (\bibinfo {year}
  {2009})%
  \bibAnnoteFile{NoStop}{zhangh2009}%
\bibitem{liuc2010}%
  \BibitemOpen
  \bibfield{author}{%
  \bibinfo {author} {\bibfnamefont{C.-X.}\ \bibnamefont{Liu}}, \bibinfo
  {author} {\bibfnamefont{X.-L.}\ \bibnamefont{Qi}}, \bibinfo {author}
  {\bibfnamefont{H.}~\bibnamefont{Zhang}}, \bibinfo {author}
  {\bibfnamefont{X.}~\bibnamefont{Dai}}, \bibinfo {author}
  {\bibfnamefont{Z.}~\bibnamefont{Fang}},\ and\ \bibinfo {author}
  {\bibfnamefont{S.-C.}\ \bibnamefont{Zhang}},\ }%
  \bibfield{journal}{%
  \bibinfo {journal} {Phys. Rev. B}\ }%
  \textbf{\bibinfo {volume} {82}},\ \bibinfo {pages} {045122} (\bibinfo {year}
  {2010})%
  \bibAnnoteFile{NoStop}{liuc2010}%
\bibitem{zhangh2013}%
  \BibitemOpen
  \bibfield{author}{%
  \bibinfo {author} {\bibfnamefont{H.}~\bibnamefont{Zhang}}, \bibinfo {author}
  {\bibfnamefont{C.-X.}\ \bibnamefont{Liu}},\ and\ \bibinfo {author}
  {\bibfnamefont{S.-C.}\ \bibnamefont{Zhang}},\ }%
  \bibfield{journal}{%
  \bibinfo {journal} {Physical review letters}\ }%
  \textbf{\bibinfo {volume} {111}},\ \bibinfo {pages} {066801} (\bibinfo {year}
  {2013})%
  \bibAnnoteFile{NoStop}{zhangh2013}%
\bibitem{teo2008}%
  \BibitemOpen
  \bibfield{author}{%
  \bibinfo {author} {\bibfnamefont{J.~C.}\ \bibnamefont{Teo}}, \bibinfo
  {author} {\bibfnamefont{L.}~\bibnamefont{Fu}},\ and\ \bibinfo {author}
  {\bibfnamefont{C.}~\bibnamefont{Kane}},\ }%
  \bibfield{journal}{%
  \bibinfo {journal} {Physical Review B}\ }%
  \textbf{\bibinfo {volume} {78}},\ \bibinfo {pages} {045426} (\bibinfo {year}
  {2008})%
  \bibAnnoteFile{NoStop}{teo2008}%
\bibitem{hsieh2012}%
  \BibitemOpen
  \bibfield{author}{%
  \bibinfo {author} {\bibfnamefont{T.~H.}\ \bibnamefont{Hsieh}}, \bibinfo
  {author} {\bibfnamefont{H.}~\bibnamefont{Lin}}, \bibinfo {author}
  {\bibfnamefont{J.}~\bibnamefont{Liu}}, \bibinfo {author}
  {\bibfnamefont{W.}~\bibnamefont{Duan}}, \bibinfo {author}
  {\bibfnamefont{A.}~\bibnamefont{Bansil}},\ and\ \bibinfo {author}
  {\bibfnamefont{L.}~\bibnamefont{Fu}},\ }%
  \bibfield{journal}{%
  \bibinfo {journal} {Nature communications}\ }%
  \textbf{\bibinfo {volume} {3}},\ \bibinfo {pages} {982} (\bibinfo {year}
  {2012})%
  \bibAnnoteFile{NoStop}{hsieh2012}%
\bibitem{takahashi2011}%
  \BibitemOpen
  \bibfield{author}{%
  \bibinfo {author} {\bibfnamefont{R.}~\bibnamefont{Takahashi}}\ and\ \bibinfo
  {author} {\bibfnamefont{S.}~\bibnamefont{Murakami}},\ }%
  \bibfield{journal}{%
  \bibinfo {journal} {Physical Review Letters}\ }%
  \textbf{\bibinfo {volume} {107}},\ \bibinfo {pages} {166805} (\bibinfo {year}
  {2011})%
  \bibAnnoteFile{NoStop}{takahashi2011}%
\bibitem{bernevig2006}%
  \BibitemOpen
  \bibfield{author}{%
  \bibinfo {author} {\bibfnamefont{B.~A.}\ \bibnamefont{Bernevig}}, \bibinfo
  {author} {\bibfnamefont{T.~L.}\ \bibnamefont{Hughes}},\ and\ \bibinfo
  {author} {\bibfnamefont{S.-C.}\ \bibnamefont{Zhang}},\ }%
  \bibfield{journal}{%
  \bibinfo {journal} {Science}\ }%
  \textbf{\bibinfo {volume} {314}},\ \bibinfo {pages} {1757} (\bibinfo {year}
  {2006})%
  \bibAnnoteFile{NoStop}{bernevig2006}%
\bibitem{moon2006}%
  \BibitemOpen
  \bibfield{author}{%
  \bibinfo {author} {\bibfnamefont{C.-Y.}\ \bibnamefont{Moon}}\ and\ \bibinfo
  {author} {\bibfnamefont{S.-H.}\ \bibnamefont{Wei}},\ }%
  \bibfield{journal}{%
  \bibinfo {journal} {Phys. Rev. B}\ }%
  \textbf{\bibinfo {volume} {74}},\ \bibinfo {pages} {045205} (\bibinfo {year}
  {2006})%
  \bibAnnoteFile{NoStop}{moon2006}%
\bibitem{svane2011}%
  \BibitemOpen
  \bibfield{author}{%
  \bibinfo {author} {\bibfnamefont{A.}~\bibnamefont{Svane}}, \bibinfo {author}
  {\bibfnamefont{N.~E.}\ \bibnamefont{Christensen}}, \bibinfo {author}
  {\bibfnamefont{M.}~\bibnamefont{Cardona}}, \bibinfo {author}
  {\bibfnamefont{A.~N.}\ \bibnamefont{Chantis}}, \bibinfo {author}
  {\bibfnamefont{M.}~\bibnamefont{van Schilfgaarde}},\ and\ \bibinfo {author}
  {\bibfnamefont{T.}~\bibnamefont{Kotani}},\ }%
  \bibfield{journal}{%
  \bibinfo {journal} {Phys. Rev. B}\ }%
  \textbf{\bibinfo {volume} {84}},\ \bibinfo {pages} {205205} (\bibinfo {year}
  {2011})%
  \bibAnnoteFile{NoStop}{svane2011}%
\bibitem{virot2013}%
  \BibitemOpen
  \bibfield{author}{%
  \bibinfo {author} {\bibfnamefont{F.~m.~c.}\ \bibnamefont{Virot}}, \bibinfo
  {author} {\bibfnamefont{R.}~\bibnamefont{Hayn}}, \bibinfo {author}
  {\bibfnamefont{M.}~\bibnamefont{Richter}},\ and\ \bibinfo {author}
  {\bibfnamefont{J.}~\bibnamefont{van~den Brink}},\ }%
  \bibfield{journal}{%
  \bibinfo {journal} {Phys. Rev. Lett.}\ }%
  \textbf{\bibinfo {volume} {111}},\ \bibinfo {pages} {146803} (\bibinfo {year}
  {2013})%
  \bibAnnoteFile{NoStop}{virot2013}%
\bibitem{carrier2004}%
  \BibitemOpen
  \bibfield{author}{%
  \bibinfo {author} {\bibfnamefont{P.}~\bibnamefont{Carrier}}\ and\ \bibinfo
  {author} {\bibfnamefont{S.-H.}\ \bibnamefont{Wei}},\ }%
  \bibfield{journal}{%
  \bibinfo {journal} {Phys. Rev. B}\ }%
  \textbf{\bibinfo {volume} {70}},\ \bibinfo {pages} {035212} (\bibinfo {year}
  {2004})%
  \bibAnnoteFile{NoStop}{carrier2004}%
\bibitem{kresse1996}%
  \BibitemOpen
  \bibfield{author}{%
  \bibinfo {author} {\bibfnamefont{G.}~\bibnamefont{Kresse}}\ and\ \bibinfo
  {author} {\bibfnamefont{J.}~\bibnamefont{Furthm{\"u}ller}},\ }%
  \bibfield{journal}{%
  \Doi{10.1103/PhysRevB.54.11169}{\bibinfo {journal} {Phys. Rev. B}}\ }%
  \textbf{\bibinfo {volume} {54}},\ \bibinfo {pages} {11169} (\bibinfo {month}
  {Oct}\ \bibinfo {year} {1996})%
  \bibAnnoteFile{NoStop}{kresse1996}%
\bibitem{mostofi2008}%
  \BibitemOpen
  \bibfield{author}{%
  \bibinfo {author} {\bibfnamefont{A.~A.}\ \bibnamefont{Mostofi}}, \bibinfo
  {author} {\bibfnamefont{J.~R.}\ \bibnamefont{Yates}}, \bibinfo {author}
  {\bibfnamefont{Y.-S.}\ \bibnamefont{Lee}}, \bibinfo {author}
  {\bibfnamefont{I.}~\bibnamefont{Souza}}, \bibinfo {author}
  {\bibfnamefont{D.}~\bibnamefont{Vanderbilt}},\ and\ \bibinfo {author}
  {\bibfnamefont{N.}~\bibnamefont{Marzari}},\ }%
  \bibfield{journal}{%
  \bibinfo {journal} {Compu. Phys. Commun}\ }%
  \textbf{\bibinfo {volume} {178}},\ \bibinfo {pages} {685} (\bibinfo {month}
  {May}\ \bibinfo {year} {2008})%
  \bibAnnoteFile{NoStop}{mostofi2008}%
\bibitem{winkler2003}%
  \BibitemOpen
  \bibfield{author}{%
  \bibinfo {author} {\bibfnamefont{R.}~\bibnamefont{Winkler}},\ }%
  \emph{\bibinfo {title} {Spin-Orbit Coupling Effects in Two-Dimensional
  Electron and Hole Systems}},\ Springer Tracts in Modern Physics\ (\bibinfo
  {publisher} {Springer},\ \bibinfo {year} {2003})%
  \bibAnnoteFile{NoStop}{winkler2003}%
\bibitem{novik2005}%
  \BibitemOpen
  \bibfield{author}{%
  \bibinfo {author} {\bibfnamefont{E.}~\bibnamefont{Novik}}, \bibinfo {author}
  {\bibfnamefont{A.}~\bibnamefont{Pfeuffer-Jeschke}}, \bibinfo {author}
  {\bibfnamefont{T.}~\bibnamefont{Jungwirth}}, \bibinfo {author}
  {\bibfnamefont{V.}~\bibnamefont{Latussek}}, \bibinfo {author}
  {\bibfnamefont{C.}~\bibnamefont{Becker}}, \bibinfo {author}
  {\bibfnamefont{G.}~\bibnamefont{Landwehr}}, \bibinfo {author}
  {\bibfnamefont{H.}~\bibnamefont{Buhmann}},\ and\ \bibinfo {author}
  {\bibfnamefont{L.}~\bibnamefont{Molenkamp}},\ }%
  \bibfield{journal}{%
  \bibinfo {journal} {Physical Review B}\ }%
  \textbf{\bibinfo {volume} {72}},\ \bibinfo {pages} {035321} (\bibinfo {year}
  {2005})%
  \bibAnnoteFile{NoStop}{novik2005}%
\bibitem{appendix}%
  \BibitemOpen
  \bibinfo {note} {Refer to Appendix for detailed form of the Kane model,
  parameters we used, mirror operators and mirror Chern number calculation.}%
  \bibAnnoteFile{Stop}{appendix}%
\bibitem{qi2006}%
  \BibitemOpen
  \bibfield{author}{%
  \bibinfo {author} {\bibfnamefont{X.-L.}\ \bibnamefont{Qi}}, \bibinfo {author}
  {\bibfnamefont{Y.-S.}\ \bibnamefont{Wu}},\ and\ \bibinfo {author}
  {\bibfnamefont{S.-C.}\ \bibnamefont{Zhang}},\ }%
  \bibfield{journal}{%
  \bibinfo {journal} {Physical Review B}\ }%
  \textbf{\bibinfo {volume} {74}},\ \bibinfo {pages} {085308} (\bibinfo {year}
  {2006})%
  \bibAnnoteFile{NoStop}{qi2006}%
\bibitem{hsieh2009a}%
  \BibitemOpen
  \bibfield{author}{%
  \bibinfo {author} {\bibfnamefont{D.}~\bibnamefont{Hsieh}}, \bibinfo {author}
  {\bibfnamefont{Y.}~\bibnamefont{Xia}}, \bibinfo {author}
  {\bibfnamefont{L.}~\bibnamefont{Wray}}, \bibinfo {author}
  {\bibfnamefont{D.}~\bibnamefont{Qian}}, \bibinfo {author}
  {\bibfnamefont{A.}~\bibnamefont{Pal}}, \bibinfo {author}
  {\bibfnamefont{J.~H.}\ \bibnamefont{Dil}}, \bibinfo {author}
  {\bibfnamefont{J.}~\bibnamefont{Osterwalder}}, \bibinfo {author}
  {\bibfnamefont{F.}~\bibnamefont{Meier}}, \bibinfo {author}
  {\bibfnamefont{G.}~\bibnamefont{Bihlmayer}}, \bibinfo {author}
  {\bibfnamefont{C.~L.}\ \bibnamefont{Kane}}, \bibinfo {author}
  {\bibfnamefont{Y.~S.}\ \bibnamefont{Hor}}, \bibinfo {author}
  {\bibfnamefont{R.~J.}\ \bibnamefont{Cava}},\ and\ \bibinfo {author}
  {\bibfnamefont{M.~Z.}\ \bibnamefont{Hasan}},\ }%
  \bibfield{journal}{%
  \bibinfo {journal} {Science}\ }%
  \textbf{\bibinfo {volume} {323}},\ \bibinfo {pages} {919} (\bibinfo {year}
  {2009})%
  \bibAnnoteFile{NoStop}{hsieh2009a}%
\bibitem{hsieh2009b}%
  \BibitemOpen
  \bibfield{author}{%
  \bibinfo {author} {\bibfnamefont{D.}~\bibnamefont{Hsieh}}, \bibinfo {author}
  {\bibfnamefont{Y.}~\bibnamefont{Xia}}, \bibinfo {author}
  {\bibfnamefont{D.}~\bibnamefont{Qian}}, \bibinfo {author}
  {\bibfnamefont{L.}~\bibnamefont{Wray}}, \bibinfo {author}
  {\bibfnamefont{J.}~\bibnamefont{Dil}}, \bibinfo {author}
  {\bibfnamefont{F.}~\bibnamefont{Meier}}, \bibinfo {author}
  {\bibfnamefont{J.}~\bibnamefont{Osterwalder}}, \bibinfo {author}
  {\bibfnamefont{L.}~\bibnamefont{Patthey}}, \bibinfo {author}
  {\bibfnamefont{J.}~\bibnamefont{Checkelsky}}, \bibinfo {author}
  {\bibfnamefont{N.}~\bibnamefont{Ong}}, \emph{et~al.},\ }%
  \bibfield{journal}{%
  \bibinfo {journal} {Nature}\ }%
  \textbf{\bibinfo {volume} {460}},\ \bibinfo {pages} {1101} (\bibinfo {year}
  {2009})%
  \bibAnnoteFile{NoStop}{hsieh2009b}%
\bibitem{souma2011}%
  \BibitemOpen
  \bibfield{author}{%
  \bibinfo {author} {\bibfnamefont{S.}~\bibnamefont{Souma}}, \bibinfo {author}
  {\bibfnamefont{K.}~\bibnamefont{Kosaka}}, \bibinfo {author}
  {\bibfnamefont{T.}~\bibnamefont{Sato}}, \bibinfo {author}
  {\bibfnamefont{M.}~\bibnamefont{Komatsu}}, \bibinfo {author}
  {\bibfnamefont{A.}~\bibnamefont{Takayama}}, \bibinfo {author}
  {\bibfnamefont{T.}~\bibnamefont{Takahashi}}, \bibinfo {author}
  {\bibfnamefont{M.}~\bibnamefont{Kriener}}, \bibinfo {author}
  {\bibfnamefont{K.}~\bibnamefont{Segawa}},\ and\ \bibinfo {author}
  {\bibfnamefont{Y.}~\bibnamefont{Ando}},\ }%
  \bibfield{journal}{%
  \bibinfo {journal} {Phys. Rev. Lett.}\ }%
  \textbf{\bibinfo {volume} {106}},\ \bibinfo {pages} {216803} (\bibinfo {year}
  {2011})%
  \bibAnnoteFile{NoStop}{souma2011}%
\bibitem{xu2011}%
  \BibitemOpen
  \bibfield{author}{%
  \bibinfo {author} {\bibfnamefont{S.-Y.}\ \bibnamefont{Xu}}, \bibinfo {author}
  {\bibfnamefont{L.}~\bibnamefont{Wray}}, \bibinfo {author}
  {\bibfnamefont{Y.}~\bibnamefont{Xia}}, \bibinfo {author}
  {\bibfnamefont{F.}~\bibnamefont{von Rohr}}, \bibinfo {author}
  {\bibfnamefont{Y.}~\bibnamefont{Hor}}, \bibinfo {author}
  {\bibfnamefont{F.}~\bibnamefont{Meier}}, \bibinfo {author}
  {\bibfnamefont{B.}~\bibnamefont{Slomski}}, \bibinfo {author}
  {\bibfnamefont{J.}~\bibnamefont{Osterwalder}}, \bibinfo {author}
  {\bibfnamefont{M.}~\bibnamefont{Neupane}}, \bibinfo {author}
  {\bibfnamefont{H.}~\bibnamefont{Lin}}, \emph{et~al.},\ }%
  \bibfield{journal}{%
  \bibinfo {journal} {arXiv preprint arXiv:1101.3985}}%
   (\bibinfo {year} {2011})%
  \bibAnnoteFile{NoStop}{xu2011}%
\bibitem{pan2011}%
  \BibitemOpen
  \bibfield{author}{%
  \bibinfo {author} {\bibfnamefont{Z.-H.}\ \bibnamefont{Pan}}, \bibinfo
  {author} {\bibfnamefont{E.}~\bibnamefont{Vescovo}}, \bibinfo {author}
  {\bibfnamefont{A.~V.}\ \bibnamefont{Fedorov}}, \bibinfo {author}
  {\bibfnamefont{D.}~\bibnamefont{Gardner}}, \bibinfo {author}
  {\bibfnamefont{Y.~S.}\ \bibnamefont{Lee}}, \bibinfo {author}
  {\bibfnamefont{S.}~\bibnamefont{Chu}}, \bibinfo {author}
  {\bibfnamefont{G.~D.}\ \bibnamefont{Gu}},\ and\ \bibinfo {author}
  {\bibfnamefont{T.}~\bibnamefont{Valla}},\ }%
  \bibfield{journal}{%
  \bibinfo {journal} {Phys. Rev. Lett.}\ }%
  \textbf{\bibinfo {volume} {106}},\ \bibinfo {pages} {257004} (\bibinfo {year}
  {2011})%
  \bibAnnoteFile{NoStop}{pan2011}%
\bibitem{jozwiak2011}%
  \BibitemOpen
  \bibfield{author}{%
  \bibinfo {author} {\bibfnamefont{C.}~\bibnamefont{Jozwiak}}, \bibinfo
  {author} {\bibfnamefont{Y.~L.}\ \bibnamefont{Chen}}, \bibinfo {author}
  {\bibfnamefont{A.~V.}\ \bibnamefont{Fedorov}}, \bibinfo {author}
  {\bibfnamefont{J.~G.}\ \bibnamefont{Analytis}}, \bibinfo {author}
  {\bibfnamefont{C.~R.}\ \bibnamefont{Rotundu}}, \bibinfo {author}
  {\bibfnamefont{A.~K.}\ \bibnamefont{Schmid}}, \bibinfo {author}
  {\bibfnamefont{J.~D.}\ \bibnamefont{Denlinger}}, \bibinfo {author}
  {\bibfnamefont{Y.-D.}\ \bibnamefont{Chuang}}, \bibinfo {author}
  {\bibfnamefont{D.-H.}\ \bibnamefont{Lee}}, \bibinfo {author}
  {\bibfnamefont{I.~R.}\ \bibnamefont{Fisher}}, \bibinfo {author}
  {\bibfnamefont{R.~J.}\ \bibnamefont{Birgeneau}}, \bibinfo {author}
  {\bibfnamefont{Z.-X.}\ \bibnamefont{Shen}}, \bibinfo {author}
  {\bibfnamefont{Z.}~\bibnamefont{Hussain}},\ and\ \bibinfo {author}
  {\bibfnamefont{A.}~\bibnamefont{Lanzara}},\ }%
  \bibfield{journal}{%
  \bibinfo {journal} {Phys. Rev. B}\ }%
  \textbf{\bibinfo {volume} {84}},\ \bibinfo {pages} {165113} (\bibinfo {year}
  {2011})%
  \bibAnnoteFile{NoStop}{jozwiak2011}%
\bibitem{cardona1963}%
  \BibitemOpen
  \bibfield{author}{%
  \bibinfo {author} {\bibfnamefont{M.}~\bibnamefont{Cardona}},\ }%
  \bibfield{journal}{%
  \bibinfo {journal} {Physical Review}\ }%
  \textbf{\bibinfo {volume} {129}},\ \bibinfo {pages} {69} (\bibinfo {year}
  {1963})%
  \bibAnnoteFile{NoStop}{cardona1963}%
\bibitem{shindo1965}%
  \BibitemOpen
  \bibfield{author}{%
  \bibinfo {author} {\bibfnamefont{K.}~\bibnamefont{Shindo}}, \bibinfo {author}
  {\bibfnamefont{A.}~\bibnamefont{Morita}},\ and\ \bibinfo {author}
  {\bibfnamefont{H.}~\bibnamefont{Kamimura}},\ }%
  \bibfield{journal}{%
  \bibinfo {journal} {Journal of the Physical Society of Japan}\ }%
  \textbf{\bibinfo {volume} {20}},\ \bibinfo {pages} {2054} (\bibinfo {year}
  {1965})%
  \bibAnnoteFile{NoStop}{shindo1965}%
\bibitem{zunger2012}%
  \BibitemOpen
  \bibfield{author}{%
  \bibinfo {author} {\bibfnamefont{J.}~\bibnamefont{Vidal}}, \bibinfo {author}
  {\bibfnamefont{X.}~\bibnamefont{Zhang}}, \bibinfo {author}
  {\bibfnamefont{V.}~\bibnamefont{Stevanovi{\'c}}}, \bibinfo {author}
  {\bibfnamefont{J.-W.}\ \bibnamefont{Luo}},\ and\ \bibinfo {author}
  {\bibfnamefont{A.}~\bibnamefont{Zunger}},\ }%
  \bibfield{journal}{%
  \bibinfo {journal} {Physical Review B}\ }%
  \textbf{\bibinfo {volume} {86}},\ \bibinfo {pages} {075316} (\bibinfo {year}
  {2012})%
  \bibAnnoteFile{NoStop}{zunger2012}%
\bibitem{chadov2010}%
  \BibitemOpen
  \bibfield{author}{%
  \bibinfo {author} {\bibfnamefont{S.}~\bibnamefont{Chadov}}, \bibinfo {author}
  {\bibfnamefont{X.~L.}\ \bibnamefont{Qi}}, \bibinfo {author}
  {\bibfnamefont{J.}~\bibnamefont{{K{\"u}bler}}}, \bibinfo {author}
  {\bibfnamefont{G.~H.}\ \bibnamefont{Fecher}}, \bibinfo {author}
  {\bibfnamefont{C.}~\bibnamefont{Felser}},\ and\ \bibinfo {author}
  {\bibfnamefont{S.~C.}\ \bibnamefont{Zhang}},\ }%
  \bibfield{journal}{%
  \bibinfo {journal} {Nature Mater.}\ }%
  \textbf{\bibinfo {volume} {9}},\ \bibinfo {pages} {541} (\bibinfo {year}
  {2010})%
  \bibAnnoteFile{NoStop}{chadov2010}%
\bibitem{lin2010}%
  \BibitemOpen
  \bibfield{author}{%
  \bibinfo {author} {\bibfnamefont{H.}~\bibnamefont{Lin}}, \bibinfo {author}
  {\bibfnamefont{L.~A.}\ \bibnamefont{Wray}}, \bibinfo {author}
  {\bibfnamefont{Y.}~\bibnamefont{Xia}}, \bibinfo {author}
  {\bibfnamefont{S.}~\bibnamefont{Xu}}, \bibinfo {author}
  {\bibfnamefont{S.}~\bibnamefont{Jia}}, \bibinfo {author}
  {\bibfnamefont{R.~J.}\ \bibnamefont{Cava}}, \bibinfo {author}
  {\bibfnamefont{A.}~\bibnamefont{Bansil}},\ and\ \bibinfo {author}
  {\bibfnamefont{M.~Z.}\ \bibnamefont{Hasan}},\ }%
  \bibfield{journal}{%
  \bibinfo {journal} {Nature Mater.}\ }%
  \textbf{\bibinfo {volume} {9}},\ \bibinfo {pages} {546} (\bibinfo {year}
  {2010})%
  \bibAnnoteFile{NoStop}{lin2010}%
\bibitem{xiao2010}%
  \BibitemOpen
  \bibfield{author}{%
  \bibinfo {author} {\bibfnamefont{D.}~\bibnamefont{Xiao}}, \bibinfo {author}
  {\bibfnamefont{Y.}~\bibnamefont{Yao}}, \bibinfo {author}
  {\bibfnamefont{W.}~\bibnamefont{Feng}}, \bibinfo {author}
  {\bibfnamefont{J.}~\bibnamefont{Wen}}, \bibinfo {author}
  {\bibfnamefont{Z.}~\bibnamefont{Wenguang}}, \bibinfo {author}
  {\bibfnamefont{X.-Q.}\ \bibnamefont{Chen}}, \bibinfo {author}
  {\bibfnamefont{G.~M.}\ \bibnamefont{Stocks}},\ and\ \bibinfo {author}
  {\bibfnamefont{Z.}~\bibnamefont{Zhang}},\ }%
  \bibfield{journal}{%
  \Doi{10.1103/PhysRevLett.105.096404}{\bibinfo {journal} {Phys. Rev. Lett.}}\
  }%
  \textbf{\bibinfo {volume} {105}},\ \bibinfo {pages} {096404} (\bibinfo
  {month} {Aug}\ \bibinfo {year} {2010})%
  \bibAnnoteFile{NoStop}{xiao2010}%
\bibitem{lin2014}%
  \BibitemOpen
  \bibfield{author}{%
  \bibinfo {author} {\bibfnamefont{S.-Y.~L.}\ \bibnamefont{Lin}}, \bibinfo
  {author} {\bibfnamefont{X.-B.}\ \bibnamefont{Yang}}, \bibinfo {author}
  {\bibfnamefont{Y.-J.}\ \bibnamefont{Zhao}}, \bibinfo {author}
  {\bibfnamefont{S.-C.}\ \bibnamefont{Wu}}, \bibinfo {author}
  {\bibfnamefont{C.}~\bibnamefont{Felser}},\ and\ \bibinfo {author}
  {\bibfnamefont{B.}~\bibnamefont{Yan}},\ }%
  \bibfield{journal}{%
  \bibinfo {journal} {to be sumibtted (arxiv number in the future)}}%
   (\bibinfo {year} {2014})%
  \bibAnnoteFile{NoStop}{lin2014}%
\bibitem{Pesin2010}%
  \BibitemOpen
  \bibfield{author}{%
  \bibinfo {author} {\bibfnamefont{D.}~\bibnamefont{Pesin}}\ and\ \bibinfo
  {author} {\bibfnamefont{L.}~\bibnamefont{Balents}},\ }%
  \bibfield{journal}{%
  \bibinfo {journal} {Nature Phys.}\ }%
  \textbf{\bibinfo {volume} {6}},\ \bibinfo {pages} {376} (\bibinfo {month}
  {Mar.}\ \bibinfo {year} {2010})%
  \bibAnnoteFile{NoStop}{Pesin2010}%
\bibitem{guo2009}%
  \BibitemOpen
  \bibfield{author}{%
  \bibinfo {author} {\bibfnamefont{H.-M.}\ \bibnamefont{Guo}}\ and\ \bibinfo
  {author} {\bibfnamefont{M.}~\bibnamefont{Franz}},\ }%
  \bibfield{journal}{%
  \bibinfo {journal} {Phys. Rev. Lett.}\ }%
  \textbf{\bibinfo {volume} {103}},\ \bibinfo {pages} {206805} (\bibinfo {year}
  {2009})%
  \bibAnnoteFile{NoStop}{guo2009}%
\bibitem{Wan2011}%
  \BibitemOpen
  \bibfield{author}{%
  \bibinfo {author} {\bibfnamefont{X.}~\bibnamefont{Wan}}, \bibinfo {author}
  {\bibfnamefont{A.~M.}\ \bibnamefont{Turner}}, \bibinfo {author}
  {\bibfnamefont{A.}~\bibnamefont{Vishwanath}},\ and\ \bibinfo {author}
  {\bibfnamefont{S.~Y.}\ \bibnamefont{Savrasov}},\ }%
  \bibfield{journal}{%
  \bibinfo {journal} {Phys. Rev. B}\ }%
  \textbf{\bibinfo {volume} {83}},\ \bibinfo {pages} {205101} (\bibinfo {year}
  {2011})%
  \bibAnnoteFile{NoStop}{Wan2011}%
\end{thebibliography}%

\begin{widetext}
\begin{appendix}
\section{Appendix}

\subsection{Kane model}

The 8-band Kane model in the bulk basis
\begin{eqnarray}
\nonumber&&\vert \Gamma_6,1/2 \rangle = \vert S \rangle \vert \uparrow \rangle \\
\nonumber&&\vert \Gamma_6,-1/2 \rangle = \vert S \rangle \vert \downarrow \rangle \\
\nonumber&&\vert \Gamma_8,3/2 \rangle = - \frac{1}{\sqrt{2}}\vert  X + iY \rangle \vert \uparrow \rangle\\
\nonumber&&\vert \Gamma_8,1/2 \rangle =  \frac{1}{\sqrt{6}}(2 \vert Z \rangle \vert \uparrow \rangle - \vert  X + iY \rangle \vert \downarrow \rangle\\
\nonumber&&\vert \Gamma_8,-1/2 \rangle =  \frac{1}{\sqrt{6}}(2 \vert Z \rangle \vert \downarrow \rangle + \vert  X - iY \rangle \vert \uparrow \rangle\\
\nonumber&&\vert \Gamma_8,-3/2 \rangle = \frac{1}{\sqrt{2}}\vert  X - iY \rangle \vert \downarrow \rangle\\
\nonumber&& \vert \Gamma_7,1/2 \rangle = -\frac{1}{\sqrt{3}} (\vert  Z \rangle \vert \uparrow \rangle + \vert X +iY \rangle \vert \downarrow \rangle\\
&& \vert \Gamma_7,-1/2 \rangle = -\frac{1}{\sqrt{3}} (- \vert  Z \rangle \vert \downarrow \rangle + \vert X - iY \rangle \vert \uparrow \rangle
\label{eqn:basis}
\end{eqnarray}

can be written as

\begin{eqnarray}
  &&H_{Kane}=\left(
	\begin{array}{cccccccc}
          T&0& -\frac{1}{\sqrt{2}}Pk_+& \sqrt{\frac{2}{3}}Pk_z& \frac{1}{\sqrt{6}} Pk_-&0&-\frac{1}{\sqrt{3}}Pk_z&-\frac{1}{\sqrt{3}}Pk_-\\
          0&T&0 &-\frac{1}{\sqrt{6}} Pk_+& \sqrt{\frac{2}{3}}Pk_z& \frac{1}{\sqrt{2}}Pk_- &-\frac{1}{\sqrt{3}}Pk_+ & \frac{1}{\sqrt{3}}Pk_z\\
	  -\frac{1}{\sqrt{2}}Pk_-&0&U+V-V_{str}&-\bar{S}_-&R&0&\frac{1}{\sqrt{2}}\bar{S}_-&-\sqrt{2}R\\
          \sqrt{\frac{2}{3}}Pk_z&-\frac{1}{\sqrt{6}} Pk_-&-\bar{S}^{\dag}_-&U-V+V_{str}&C&R&\sqrt{2}V&-\sqrt{\frac{3}{2}}\tilde{S}_-\\
	  \frac{1}{\sqrt{6}} Pk_+&\sqrt{\frac{2}{3}}Pk_z&R^{\dag}&C^{\dag}&U-V+V_{str}&\bar{S}^{\dag}_+&-\sqrt{\frac{3}{2}\tilde{S}_+}&-\sqrt{2}V\\
	  0&\frac{1}{\sqrt{2}}Pk_+&0&R^{\dag}&\bar{S}_+&U+V-V_{str}&\sqrt{2}R^\dag&\frac{1}{\sqrt{2}}\bar{S}_+\\
          -\frac{1}{\sqrt{3}}Pk_z&-\frac{1}{\sqrt{3}}Pk_-&\frac{1}{\sqrt{2}}\bar{S}^\dag_-&\sqrt{2}V&-\sqrt{\frac{3}{2}}\tilde{S}^\dag_+&\sqrt{2}R&U-\Delta&C\\
          -\frac{1}{\sqrt{3}}Pk_+&\frac{1}{\sqrt{3}}Pk_z&-\sqrt{2}^\dag&-\sqrt{\frac{3}{2}}\tilde{S}^\dag_-&-\sqrt{2}V&\frac{1}{\sqrt{2}}\bar{S}^\dag_+&C^\dag&U-\Delta
	\end{array}
\label{eq:ham_kane}
	\right)
\end{eqnarray}
where
\begin{eqnarray}
\nonumber&&T = E_c(z) + \frac{\hbar^2}{2m_0}[(2F+1)k^2_{||}+k_z(2F+1)k_z]\\
\nonumber&&U = E_v(z) - \frac{\hbar^2}{2m_0}(\gamma_1 k^2_{||} + k_z \gamma_1 k_z)\\
\nonumber&&V = - \frac{\hbar^2}{2m_0}(\gamma_2 k^2_{||} - 2 k_z \gamma_2 k_z)\\
\nonumber&&R = - \frac{\hbar^2}{2m_0}(\sqrt{3} \mu k^2_+ - \sqrt{3} \bar{\gamma}k^2_-)\\
\nonumber&&\bar{S}_{\pm} = - \frac{\hbar^2}{2m_0} \sqrt{3}k_{\pm}(\{\gamma_3,k_z\} + [\kappa,k_z])\\
\nonumber&&\tilde{S}_\pm = - \frac{\hbar^2}{2m_0} \sqrt{3}k_{\pm}(\{\gamma_3,k_z\} -\frac{1}{3} [\kappa,k_z])\\
\nonumber&&C =  \frac{\hbar^2}{2m_0}k_-[\kappa,k_z]\\
&&V_{str} = a \epsilon_{xx}
\end{eqnarray}
Here $\gamma_1$, $\gamma_2$, $\gamma_3$, $\bar{\gamma} = (\gamma_2 + \gamma_3)/2$ and $\mu = (\gamma_3 - \gamma_2)/2$ are parameters depending on materials; $a$ is related to deformation potentials and $\epsilon_{xx}$ is the strain tensor; $\{$,$\}$ and $[,]$ are anticommutative and commutative operators. The bulk inversion asymmetrical Hamiltonian is expressed as

\begin{eqnarray}
  &&H_{BIA}=\left(
	\begin{array}{cccccccc}
         0&0&0&0&0&0&0&0\\
         0&0&0&0&0&0&0&0\\
         0&0&0&-\frac{1}{2}C_kk_-&C_kk_z&-\frac{\sqrt{3}}{2}C_kk_-&\frac{1}{2\sqrt{2}}C_kk_+&\frac{1}{\sqrt{2}}C_kk_z\\
         0&0&-\frac{1}{2}C_kk_-&0&\frac{\sqrt{3}}{2}C_kk_+&-C_kk_z&0&-\frac{\sqrt{3}}{2\sqrt{2}}C_kk_+\\
         0&0&C_kk_z&\frac{\sqrt{3}}{2}C_kk_-&0&-\frac{1}{2}C_kk_+&\frac{\sqrt{3}}{2\sqrt{2}}C_kk_-&0\\
         0&0&-\frac{\sqrt{3}}{2}C_kk_+&-C_kk_z&-\frac{1}{2}C_kk_-&0&\frac{1}{\sqrt{2}}C_kk_z&-\frac{1}{2\sqrt{2}}C_kk_-\\
         0&0&\frac{1}{2\sqrt{2}}C_kk_-&0&\frac{\sqrt{3}}{2\sqrt{2}}C_kk_+&\frac{1}{\sqrt{2}}C_kk_z&0&0\\
         0&0&\frac{1}{\sqrt{2}}C_kk_z&-\frac{\sqrt{3}}{2\sqrt{2}}C_kk_-&0&-\frac{1}{2\sqrt{2}}C_kk_+&0&0
	\end{array}
\label{eq:ham_bia8}
	\right)
\end{eqnarray}
where $C_k$ depends on materials.

\subsection{Parameters for the Kane model}
The parameters for the 8-band Kane model are listed in Table \ref{tab1} for HgS and HgTe. The strain we considered in this paper is along x direction. The valence band offset between HgS and HgTe layers is about 90 meV. The parameters for HgS are obtained by fitting the results of \textit{ab-initio} calculation while parameters for HgTe are taken from Ref. \onlinecite{novik2005}. Without changing topological nature, we adjust $E_g = -0.503$ eV and $V_{str} = -0.4 $ eV for HgTe.

\begin{center}
\begin{table}[htb]
  \centering
  \begin{minipage}[t]{1.\linewidth}
	  \caption{ The parameters of Kane model for HgS and HgTe.  }
\label{tab1}
\hspace{-1cm}
\begin{tabular}
[c]{ccccccccccc}\hline\hline
 &$E_g$[eV]&$\Delta_{so}$[eV]&P[$eV \cdot$ \AA]& $\gamma_1$&$\gamma_2$&$\gamma_3$&$\kappa$&$V_{str}$[eV]&F&$C_k$\\
HgS&-0.6042& -0.1078 &1.2&0.3&0.08&0.01&0&0&0&-0.30 \\
HgTe&-0.503&1.08&8.46 & 4.1 &0.5&1.3&-0.4&-0.4&0&-0.19 \\\hline\hline\hline
\label{para_kane}
\end{tabular}
  \end{minipage}
\end{table}
\end{center}

\subsection{Mirror and spin operator along z direction}
In the coordinate system described in the main text, the Kane model is invariant under the mirror operation $\mathcal{M}_{z}$ along the $z$ direction ($\mathcal{M}_{z} H_{Kane}(k_x, k_y, k_z) \mathcal{M}^{-1}_{z} = H_{Kane}(k_x, k_y, -k_z)$). The mirror operator with respect to $k_z = 0$ plane is denoted as $\mathcal{M}_z = C_2 \otimes \emph{i}$, where $C_2$ is the two fold rotation operator along z axis and $\emph{i}$ is the inversion operator. The mirror operator $\mathcal{M}_z$ in the basis expressed in Eq. \ref{eqn:basis} can be expressed as
\begin{eqnarray}
 &&\mathcal{M}_z=\left(
	\begin{array}{cccccccc}
          i&0&0&0&0&0&0&0\\
          0&-i&0&0&0&0&0&0\\
          0&0&i&0&0&0&0&0\\
          0&0&0&-i&0&0&0&0\\
          0&0&0&0&i&0&0&0\\
          0&0&0&0&0&-i&0&0\\
          0&0&0&0&0&0&-i&0\\
          0&0&0&0&0&0&0&i
    \end{array}
\label{eq:mirror_z}
	\right)
\end{eqnarray}
The mirror parity $\pm i$ for each basis could be viewed in this way. Take $\vert \Gamma_7,-\frac{1}{2}\rangle = -\frac{1}{\sqrt{3}} (- \vert  Z \rangle \vert \downarrow \rangle + \vert X - iY \rangle \vert \uparrow \rangle$ for example. Under mirror symmetry operation $M_z = C_2 \otimes \emph{i} = e^{i*\pi/2*\sigma_z}\otimes \emph{i} = i\sigma_z \otimes \emph{i}$, $\vert \uparrow \rangle$ is transformed to $+i \vert \uparrow \rangle$, $\vert \downarrow \rangle$ $\rightarrow$  $-i \vert \downarrow \rangle$, X(Y) $\rightarrow$ X(Y), Z $\rightarrow$ -Z. All these transformations lead $\vert \Gamma_7,-\frac{1}{2}\rangle$ to $+i \vert \Gamma_7,-\frac{1}{2}\rangle$. Thus, the mirror operator matrix element in the basis $\vert \Gamma_7,-\frac{1}{2}\rangle$ would be $+i$. In a similar way, one can show that the states $\vert \Gamma_6,\frac{1}{2}\rangle$, $\vert \Gamma_8,\frac{3}{2}\rangle$ and $\vert \Gamma_8,-\frac{1}{2}\rangle$ have mirror parity $+i$, while the other basis have mirror parity $-i$.

The spin operator along the z direction reads

\begin{eqnarray}
 &&S_z=\frac{\hbar}{2}\left(
	\begin{array}{cccccccc}
        1&0&0&0&0&0&0&0\\
        0&-1&0&0&0&0&0&0\\
        0&0&1&0&0&0&0&0\\
        0&0&0&1/3&0&0&-2\sqrt{2}/3&0\\
        0&0&0&0&-1/3&0&0&-2\sqrt{2}/3\\
        0&0&0&0&0&-1&0&0\\
        0&0&-2\sqrt{2}/3&0&0&0&-1/3&0\\
        0&0&0&-2\sqrt{2}/3&0&0&0&1/3
    \end{array}
\label{eq:s_z}
	\right)
\end{eqnarray}

For a state $\Psi$, spin orientation is calculated by $<\Psi|S_z|\Psi>$. In the main text, we have shown that spin orientation always follows mirror parity for both HgTe and HgS. This is true for any system described by the Kane model, as shown in Fig. 3(a), in which orbital-resolved (s or p orbital) spin textures are the same as the total spin texture. In a general case, orbital-resolved spin textures might be different for different orbitals. In this case, one can easily see that s-orbital-resolved spin texture should follow mirror parity since no orbital angular momentum is involved for s orbital. For the p orbital, mirror parity is determined by the combination of spin and orbital components. Therefore, one can use s-orbital-resolved spin texture to determine mirror parity.

\subsection{Mirror Chern number}
The original effective four band model for HgTe on $k_x-k_y$ plane, which is expressed in the basis of $\vert \Gamma_6, \pm 1/2\rangle$ and $\vert \Gamma_8, \mp 3/2\rangle$, can be written as
\begin{eqnarray}
  &&H_{HgTe}=\left(
	\begin{array}{cccc}
          T&0& -\frac{1}{\sqrt{2}}Pk_+& 0 \\
          0&T&0&\frac{1}{\sqrt{2}} Pk_-\\
          -\frac{1}{\sqrt{2}}Pk_-&0&U-V+str&0\\
	  0&\frac{1}{\sqrt{2}} Pk_+&0&U-V+V_{str}
    \end{array}
\label{eq:Ap-HgTe}
	\right)
\end{eqnarray}

The mirror operator with respect to z direction in above basis takes form
\begin{eqnarray}
  &&\mathcal{M}_z=\left(
	\begin{array}{cccc}
          i&0&0&0\\
          0&-i&0&0\\
          0&0&i&0\\
          0&0&0&-i
    \end{array}
\label{eq:Ap-mHgTe}
	\right)
\end{eqnarray}
Since the mirror operator commutes with the 4-band effective Hamiltonian(with $k_z = 0$), we could diagonalize the 4-band effective Hamiltonian of HgTe, as a consequence, by selecting the following common eigenvectors: $\frac{1}{2}(1,0,0,0)^T$ and $\frac{1}{2}(0,0,1,0)^T$ with mirror eigenvalue $i$; $\frac{1}{2}(0,1,0,0)^T$ and $\frac{1}{2}(0,0,0,1)^T$ with mirror eigenvalue $-i$. In the basis above, the effective Hamiltonian of HgTe can be expressed in a block diagonal matrix, which reads

\begin{eqnarray}
&&H_{HgTe}=\left(
	\begin{array}{cccc}
          T&-\frac{1}{\sqrt{2}}Pk_+& 0&0 \\
	  -\frac{1}{\sqrt{2}}Pk_-&U-V+V_{str}&0&\\
          0&0&T&\frac{1}{\sqrt{2}} Pk_-\\
	  0&0&\frac{1}{\sqrt{2}} Pk_+&U-V+V_{str}
    \end{array}
\label{eq:Ap-bHgTe}
	\right)
\end{eqnarray}

The first block owns mirror eigenvalue $+i$ while the second block has mirror eigenvalue $-i$. The block Hamiltonian could be further written in a more compact way: $H_{4,HgTe} = \frac{T+\bar{U}}{2} \tau_0 \otimes \sigma_0 + \frac{T - \bar{U}}{2} \tau_z \otimes \sigma_0 - \frac{1}{\sqrt{2}}Pk_x \tau_z \otimes \sigma_x + \frac{1}{\sqrt{2}} Pk_y \tau_0 \otimes \sigma_y$, where $\bar{U} \equiv U - V + V_{str}$. One could determine the chirality by checking the winding number for each block\cite{qi2006},
\begin{eqnarray}
&&n_{xy} = \int_{BZ} \frac{d^2k}{4\pi} \hat{\mathbf{d}} \cdot (\frac{\partial \hat{\mathbf{d}}}{\partial k_x} \times \frac{\partial \hat{\mathbf{d}}}{\partial k_y})
\end{eqnarray}
where $\hat{\mathbf{d}} = \frac{1}{|\mathbf{d}|}(d_x, d_y, d_z)$ and $\hat{\mathbf{d}} = (-(+)\frac{1}{\sqrt{2}} Pk_x, -\frac{1}{\sqrt{2}}Pk_y, +(-)\frac{T-\bar{U}}{2})$ for fist(second) block. We obtain that the winding number of the first block is $n_{+i} = -1$ while the winding number for the second block is $n_{-i} = 1$. Thus, the mirror Chern number for HgTe is $n_{M}(HgTe) \equiv \frac{1}{2}(n_{+i} - n_{-i}) = -1$.

Similarly, an effective 4-band model, in the basis of $\vert \Gamma_6 \pm 1/2\rangle$ and $\vert \Gamma_7 \pm 1/2\rangle$, can well describe the bulk property of HgS. It takes form
\begin{eqnarray}
  &&H_{HgS}=\left(
	\begin{array}{cccc}
          T&0&-\frac{1}{\sqrt{3}}Pk_z&-\frac{1}{\sqrt{3}} Pk_-\\
          0&T&-\frac{1}{\sqrt{3}} Pk_+& \frac{1}{\sqrt{3}}Pk_z\\
          -\frac{1}{\sqrt{3}}Pk_z&-\frac{1}{\sqrt{3}} Pk_-&U-\Delta_{so}&0\\
          -\frac{1}{\sqrt{3}}Pk_+&\frac{1}{\sqrt{3}}Pk_z&0&U-\Delta_{so}
    \end{array}
\label{eq:Ap-HgS}
	\right)
\end{eqnarray}

The mirror operator with respect to z direction in the above basis is expressed as
\begin{eqnarray}
  &&\mathcal{M}_z=\left(
	\begin{array}{cccc}
          i&0&0&0\\
          0&-i&0&0\\
          0&0&-i&0\\
          0&0&0&i
    \end{array}
\label{eq:Ap-mHgS}
	\right)
\end{eqnarray}
Similarly, we could diagonalize the 4-band effective Hamiltonian of HgS for $k_x-k_y$ plane in momentum space by selecting the following common eigenvectors: $\frac{1}{2}(1,0,0,0)^T$ and $\frac{1}{2}(0,0,0,1)^T$ with mirror eigenvalue $i$; $\frac{1}{2}(0,1,0,0)^T$ and $\frac{1}{2}(0,0,1,0)^T$ with mirror eigenvalue $-i$. In the basis above, the effective Hamiltonian of HgS can be expressed in a block diagonal matrix, which reads

\begin{eqnarray}
&&H_{HgS}=\left(
	\begin{array}{cccc}
          T&-\frac{1}{\sqrt{3}}Pk_-& 0&0 \\
          -\frac{1}{\sqrt{3}}Pk_+&U-\Delta_{so}&0&0\\
          0&0&T&-\frac{1}{\sqrt{3}} Pk_+\\
          0&0&-\frac{1}{\sqrt{3}} Pk_-&U-\Delta_{so}
    \end{array}
\label{eq:Ap-bHgS}
	\right)
\end{eqnarray}

By a similar calculation, the winding number of the first block turns out to be $n_{+i} = 1$ while the winding number for the second block is $n_{-i} = -1$. Thus, the mirror Chern number for HgS is $n_{M}(HgS) \equiv \frac{1}{2}(n_{+i} - n_{-i}) = 1$.

\end{appendix}

\end{widetext}
\end{document}